\documentclass[aps,prd,print,groupedaddress,nofootinbib,eqsecnum,onecolumn]{revtex4}
\usepackage{eurosym}
\usepackage{amsmath}
\usepackage{amsthm}
\usepackage{amssymb}
\usepackage{slashed}
\usepackage{hyperref}
\usepackage{epstopdf}
\usepackage[english]{babel}
\usepackage[latin1]{inputenc}
\usepackage{amsmath, amsthm, amssymb}
\usepackage[pdftex]{graphicx}
\usepackage{amssymb}
\usepackage{latexsym}
\usepackage{amstext}
\usepackage{epstopdf}
\usepackage{floatflt}
\usepackage{hyperref}
\usepackage{enumitem}
\usepackage{multirow}
\usepackage{pdflscape}
\usepackage{makecell}
\usepackage{nopageno}

\numberwithin{equation}{section}
\begin{document}

\begin{titlepage}
\begin{center}

\vskip 3.0cm

{\bf \huge Vinberg's T-Algebras}\\
\vskip 1.0cm
{\bf \Large From Exceptional Periodicity to Black Hole Entropy}

\vskip 3.0cm

{\bf \large Alessio Marrani}

\vskip 40pt

{\it Instituto de F\'{\i}sica Teorica, Dep.to de F\'{\i}sica,\\
Universidad de Murcia, Campus de Espinardo, E-30100, Spain}\\
\texttt{alessio.marrani@um.es},

\vskip 5pt

\vskip 40pt


\vskip 40pt


\vskip 30pt



\end{center}

\vskip 95pt

\begin{center} {\bf ABSTRACT}\\[3ex]\end{center}

We introduce the so-called Magic Star (MS) projection within the root
lattice of finite-dimensional exceptional Lie algebras, and relate it to
rank-3 simple and semi-simple Jordan algebras. By relying on the Bott
periodicity of reality and conjugacy properties of spinor representations,
we present the so-called Exceptional Periodicity (EP) algebras, which are
finite-dimensional algebras, violating the Jacobi identity, and providing an
alternative with respect to Kac-Moody infinite-dimensional Lie algebras.
Remarkably, also EP algebras can be characterized in terms of a MS\
projection, exploiting special Vinberg T-algebras, a class of generalized
Hermitian matrix algebras introduced by Vinberg in the '60s within his
theory of homogeneous convex cones. As physical applications, we highlight
the role of the invariant norm of special Vinberg T-algebras in
Maxwell-Einstein-scalar theories in 5 space-time dimensions, in which the
Bekenstein-Hawking entropy of extremal black strings can be expressed in
terms of the cubic polynomial norm of the T-algebras.

\vskip 65pt

\begin{center}
Parallel talk presented at the \textit{34th International Colloquium on Group Theoretical Methods in Physics},\\Strasbourg, July 18-22, 2022
\end{center}





%
\vfill

\end{titlepage}

\newpage \setcounter{page}{1} \numberwithin{equation}{section}

\section{Projecting root lattices onto the Magic Star}

\begin{figure}[h]
\centering
\includegraphics[width=0.60\textwidth]{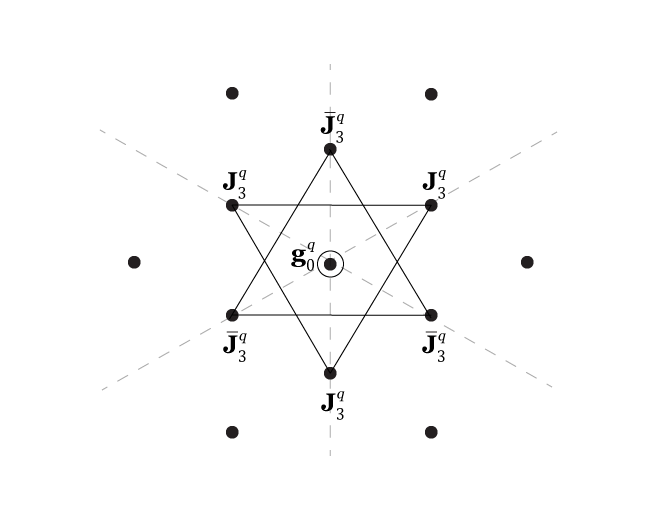}
\caption{The Magic Star of exceptional Lie algebras \protect\cite{mukai,pt1}%
. $\mathbf{J}_{3}^{q}$ denotes a rank-3 simple Jordan algebra, realized as
matrix algebra of $3\times 3$ Hermitian matrices over Hurwitz's division
algebras $\mathbb{A}=\mathbb{R},\mathbb{C},\mathbb{H},\mathbb{O}$ (of real
dimension $q=\dim _{\mathbb{R}}\mathbb{A}=1,2,4,8$, resp.). The limit case
of $\mathfrak{g}_{2}$ (corresponding to $q=-2/3$) corresponds to a trivial
Jordan algebra, given by the identity element only: $\mathbf{J}%
_{3}^{-2/3}\equiv \mathbb{I}:=diag(1,1,1)$.}
\label{fig:MagicStar}
\end{figure}

Within the $r$-dimensional root lattice of $\mathfrak{g}_{2}$, $\mathfrak{f}%
_{4}$, $\mathfrak{e}_{6}$, $\mathfrak{e}_{7}$ and $\mathfrak{e}_{8}$ (with $%
r=2,4,6,7,8$, resp.), one can find a plane (defined by the two Cartans of an
$\mathfrak{a}_{2}$ subalgebra) on which the projection of the roots results
into the so-called \textquotedblleft Magic Star" (MS)\textit{\ }(reported in
Fig. \ref{fig:MagicStar}). To the best of our knowledge, the MS was firstly
observed in late '90s by Mukai\footnote{%
Mukai used the name \textit{\textquotedblleft }$\mathfrak{g}_{2}$ \textit{%
decomposition"}.} \cite{mukai}, and later re-discovered and treated in some
detail by Truini \cite{pt1} (see also \cite{pt2}), within a different
approach relying Jordan Pairs \cite{loos}; see also \cite{3bis}.

The existence of the MS relies on the so-called (not necessarily maximal,
generally non-symmetric) MS embedding/decomposition\footnote{%
For an application to supergravity, see \cite{FMZ-D=5} (where MS embedding
was named Jordan pairs' embedding), as well as \cite{super-Ehlers}, in which
the MS embedding was elucidated to be nothing but the $D=5$ instance of the
so-called super-Ehlers embedding.}%
\begin{equation}
\mathfrak{qconf}\left( \mathbf{J}_{3}^{q}\right) \supset \mathfrak{a}_{%
\mathbf{2}}\oplus \mathfrak{str}_{0}\left( \mathbf{J}_{3}^{q}\right) ,
\label{1}
\end{equation}%
where $\mathfrak{qconf}\left( \mathbf{J}_{3}^{q}\right) $ and $\mathfrak{str}%
_{0}\left( \mathbf{J}_{3}^{q}\right) $ stand for the \textit{quasi-conformal}
resp. the \textit{reduced structure} Lie algebra of $\mathbf{J}_{3}^{q}$
(see e.g. \cite{GKN,GP-04} for basic definitions, and a list of Refs.).

Over $\mathbb{C}$, (\ref{1}) implies \cite{pt1,pt2}%
\begin{equation}
\mathfrak{qconf}\left( \mathbf{J}_{3}^{q}\right) =\mathfrak{a}_{\mathbf{2}%
}\oplus \mathfrak{str}_{0}\left( \mathbf{J}_{3}^{q}\right) \oplus \mathbf{3}%
\times \mathbf{J}_{3}^{q}\oplus \overline{\mathbf{3}}\times \overline{%
\mathbf{J}_{3}^{q}}.  \label{2}
\end{equation}%
Upon setting $q=8,4,2,1,0,-2/3,-1$, one obtains the exceptional sequence (or
exceptional series) \ref{grouptheorytable}, cf. e.g. \cite{LM-1}\footnote{%
Note that we consider $\mathfrak{b}_{3}$, corresponding to $q=-1/3$ and
absent in \cite{LM-1}.}).

\begin{table}[h]
\begin{center}
\begin{tabular}{|c||c|c|c|c|c|c|c|c|}
\hline
$q$ & $8$ & $4$ & $2$ & $1$ & $0$ & $-1/3$ & $-2/3$ & $-1$ \\ \hline
\rule[-1mm]{0mm}{6mm} $\mathfrak{qconf}\left( \mathbf{J}_{3}^{q}\right) $ & $%
\mathfrak{e}_{8}$ & $\mathfrak{e}_{7}$ & $\mathfrak{e}_{6}$ & $\mathfrak{f}%
_{4}$ & $\mathfrak{d}_{4}$ & $\mathfrak{b}_{3}$ & $\mathfrak{g}_{2}$ & $%
\mathfrak{a}_{2}$ \\ \hline
\rule[-1mm]{0mm}{6mm} $\mathfrak{str}_{0}\left( \mathbf{J}_{3}^{q}\right)$ & $\mathfrak{e}%
_{6}$ & $\mathfrak{a}_{5}$ & $\mathfrak{a}_{2}\oplus \mathfrak{a}_{2}$ & $%
\mathfrak{a}_{2}$ & $\mathbb{C}\oplus \mathbb{C}$ & $\mathbb{C}$ & $%
0 $ & $-$ \\ \hline
\end{tabular}%
\end{center}
\par
\label{grouptheorytable}
\end{table}
$\mathbf{J}_{3}^{q}$ stands for the rank-$3$ simple Jordan algebra \cite%
{JVNW} (cfr. e.g. \cite{mc}, and Refs. therein) associated to the parameter $%
q$, which for $q=8,4,2,1$ is the real dimension of the division algebra $%
\mathbb{A}$ on which the corresponding Jordan algebra is realized as a $%
3\times 3$ generalized matrix algebra with the property of $\mathbb{A}$%
-Hermiticity: $q=\dim _{\mathbb{R}}\mathbb{A}=8,4,2,1$ for $\mathbb{A}=%
\mathbb{O},\mathbb{H},\mathbb{C},\mathbb{R}$, resp., and $\mathbf{J}%
_{3}^{q}\equiv \mathbf{J}_{3}^{\mathbb{A}}\equiv H_{3}\left( \mathbb{A}%
\right) $ are equivalent notations. Remarkably, $\mathfrak{qconf}\left(
\mathbf{J}_{3}^{q}\right) $ and $\mathfrak{str}_{0}\left( \mathbf{J}%
_{3}^{q}\right) $ span the entries of the fourth resp. second row/column of
the Freudenthal-Tits Magic Square \cite{tits1,freu} when setting $q=8,4,2,1$%
. From the classification of finite-dimensional, semi-simple cubic Jordan
algebras \cite{JVNW}, $\mathbf{J}_{3}^{0}\equiv \mathbb{C}\oplus \mathbb{C}%
\oplus \mathbb{C}$ is the completely factorized (triality symmetric) rank-3
Jordan algebra, whereas $\mathbf{J}_{3}^{-1/3}\equiv \mathbb{C}\oplus
\mathbb{C}$ and $\mathbf{J}_{3}^{-2/3}\equiv \mathbb{C}$ are its partial and
total diagonal degenerations, respectively.

Within this report, we will consider things over $\mathbb{R}$. In this case,
there are \textit{at least }two non-compact real forms of the
\textquotedblleft enlarged" exceptional sequence $\left\{ \mathfrak{qconf}%
\left( \mathbf{J}_{3}^{q}\right) \right\} _{q=8,4,2,1,0,-1/3,-2/3,-1}$ which
can be easily interpreted in terms of symmetries of rank-3 real Jordan
algebras : they are given in Tables \ref{grouptheorytable2} and Table \ref%
{grouptheorytable3}. and they both pertain to the following non-compact,
real form of (\ref{2})) : $\mathfrak{qconf}$ $\mathfrak{e}_{8}$

\begin{equation}
\mathfrak{qconf}\left( \mathbf{J}_{3}^{q}\right) =\mathfrak{sl}_{3,\mathbb{R}%
}\oplus \mathfrak{str}_{0}\left( \mathbf{J}_{3}^{q}\right) \oplus \mathbf{3}%
\times \mathbf{J}_{3}^{q}\oplus \mathbf{3}^{\prime }\times \mathbf{J}%
_{3}^{q\prime }.  \label{2tris}
\end{equation}

\begin{table}[h]
\begin{center}
\begin{tabular}{|c||c|c|c|c|c|c|c|c|}
\hline
$q$ & $8$ & $4$ & $2$ & $1$ & $0$ & $-1/3$ & $-2/3$ & $-1$ \\ \hline
\rule[-1mm]{0mm}{6mm} $\mathfrak{qconf}\left( \mathbf{J}_{3}^{q}\right) $ & $%
\mathfrak{e}_{8(8)}$ & $\mathfrak{e}_{7(7)}$ & $\mathfrak{e}_{6(6)}$ & $%
\mathfrak{f}_{4(4)}$ & $\mathfrak{so}_{4,4}$ & $\mathfrak{so}_{4,3}$ & $%
\mathfrak{g}_{2(2)}$ & $\mathfrak{sl}_{3,\mathbb{R}}$ \\ \hline
\rule[-1mm]{0mm}{6mm} $\mathfrak{str}_{0}(\mathbf{J}_{3}^{q})$ & $\mathfrak{e%
}_{6(6)}$ & $\mathfrak{sl}_{6,\mathbb{R}}$ & $\mathfrak{sl}_{3,\mathbb{R}%
}\oplus \mathfrak{sl}_{3,\mathbb{R}}$ & $\mathfrak{sl}_{3,\mathbb{R}}$ & $%
\mathbb{R}\oplus \mathbb{R}$ & $\mathbb{R}$ & $0$ & $-$ \\ \hline
\end{tabular}%
\end{center}
\caption{The split real form of the exceptional sequence. In this case, for $%
q=8,4,2,1$, $\mathbf{J}_{3}^{q}\equiv \mathbf{J}_{3}^{\mathbb{A}_{s}}\equiv
H_{3}(\mathbb{A}_{s})$, where $\mathbb{A}_{s}$ is the split form of $\mathbb{%
A}=\mathbb{O},\mathbb{H},\mathbb{C}$, respectively. }
\label{grouptheorytable2}
\end{table}

\begin{table}[h]
\begin{center}
\begin{tabular}{|c||c|c|c|c|c|c|c|c|}
\hline
$q$ & $8$ & $4$ & $2$ & $1$ & $0$ & $-1/3$ & $-2/3$ & $-1$ \\ \hline
\rule[-1mm]{0mm}{6mm} $\mathfrak{qconf}\left( \mathbf{J}_{3}^{q}\right) $ & $%
\mathfrak{e}_{8(-24)}$ & $\mathfrak{e}_{7(-5)}$ & $\mathfrak{e}_{6(2)}$ & $%
\mathfrak{f}_{4(4)}$ & $\mathfrak{so}_{4,4}$ & $\mathfrak{so}_{4,3}$ & $%
\mathfrak{g}_{2(2)}$ & $\mathfrak{sl}_{3,\mathbb{R}}$ \\ \hline
\rule[-1mm]{0mm}{6mm} $\mathfrak{str}_{0}(\mathbf{J}_{3}^{q})$ & $\mathfrak{e%
}_{6(-26)}$ & $\mathfrak{su}_{6}^{\ast }$ & $(\mathfrak{sl}_{3,\mathbb{C}})_{%
\mathbb{R}}$ & $\mathfrak{sl}_{3,\mathbb{R}}$ & $\mathbb{R}\oplus \mathbb{R}$
& $\mathbb{R}$ & $0$ & $-$ \\ \hline
\end{tabular}%
\end{center}
\caption{Another (non-split) non-compact real form of the exceptional
sequence}
\label{grouptheorytable3}
\end{table}

\section{Spinor content of exceptional Lie algebras and Fierz identities in $%
8+q$ dimensions}

The following maximal, Jordan algebraic embeddings
\begin{eqnarray}
\mathbf{J}_{3}^{\mathbb{A}} &\supset &\mathbb{R}\oplus \mathbf{J}_{2}^{%
\mathbb{A}},  \notag \\
\mathbf{J}_{3}^{\mathbb{A}_{s}} &\supset &\mathbb{R}\oplus \mathbf{J}_{2}^{%
\mathbb{A}_{s}},  \label{A}
\end{eqnarray}%
enjoy the following matrix realization as ($r_{i}\in \mathbb{R}$, $A_{i}\in
\mathbb{A}$ or $\mathbb{A}_{s}$, $i=1,2,3$)%
\begin{equation}
\mathbf{J}_{3}^{\mathbb{A}}\ni J=\left(
\begin{array}{ccc}
r_{1} & A_{1} & \overline{A}_{2} \\
\overline{A}_{1} & r_{2} & A_{3} \\
A_{2} & \overline{A}_{3} & r_{3}%
\end{array}%
\right) \Rightarrow J^{\prime }=\left(
\begin{array}{ccc}
r_{1} & A_{1} & 0 \\
\overline{A}_{1} & r_{2} & 0 \\
0 & 0 & r_{3}%
\end{array}%
\right) \in \mathbb{R}\oplus \mathbf{J}_{2}^{\mathbb{A}},
\end{equation}%
where the bar denotes the conjugation in $\mathbb{A}$ or in $\mathbb{A}_{s}$%
. Usually, the matrix elements $r_{1}$ and $r_{2}$ are associated to
lightcone degrees of freedom, i.e.%
\begin{equation}
r_{1}:=x_{+}+x_{-},~r_{2}:=x_{+}-x_{-}.  \label{lightcone}
\end{equation}%
Furthermore, the following algebraic isomorphisms hold (cf. e.g. \cite%
{29-of-EYM}) :%
\begin{eqnarray}
\mathbf{J}_{2}^{\mathbb{A}} &\sim &\mathbf{\Gamma }_{1,q+1}; \\
\mathbf{J}_{2}^{\mathbb{A}_{s}} &\sim &\mathbf{\Gamma }_{q/2+1,q/2+1},
\end{eqnarray}%
where $\mathbf{\Gamma }_{1,q+1}$ and $\mathbf{\Gamma }_{q/2+1,q/2+1}$ are
(generally simple) Jordan algebras of rank 2 with a quadratic form of
(Lorentian resp. Kleinian) signature $\left( 1,q+1\right) $ resp. $%
(q/2+1,q/2+1$), i.e. the Clifford algebras of $O\left( 1,q+1\right) $ resp. $%
O(q/2+1,q/2+1)$; for this reason, it is customary to refer to (\ref{A}) as
to the the \textit{spin-factor }embeddings.

By setting $\mathbb{A}=\mathbb{O}$, i.e. $q=8$, in (\ref{A}), and
considering the various symmetries of Jordan algebras, one obtains the
graded structure of suitable real forms of finite-dimensional exceptional
Lie algebras with respect to the corresponding pseudo-orthogonal Lie
algebras, thus obtaining the spinor content of the exceptional algebras
themselves :

\begin{enumerate}
\item For what concerns the derivations $\mathfrak{der}$ (namely, the Lie
algebra of the automorphism group) of the rank-3 Jordan algebras, one
obtains the 2-graded structure of the real, compact form of $\mathfrak{f}%
_{4} $, namely :
\begin{equation}
\mathfrak{der}\left( \mathbf{J}_{3}^{\mathbb{O}}\right) \supset ^{m,s}%
\mathfrak{der}\left( \mathbb{R}\oplus \mathbf{J}_{2}^{\mathbb{O}}\right)
\Leftrightarrow \left\{
\begin{array}{c}
\mathfrak{f}_{4(-52)}\supset ^{m,s}\mathfrak{so}_{9}; \\
~ \\
\mathfrak{f}_{4(-52)}=\mathfrak{so}_{9}\oplus \mathbf{16},%
\end{array}%
\right.  \label{f4}
\end{equation}%
where $\mathbf{16}$ is the Majorana spinor irrepr. of $\mathbf{so}_{9}$, and
the upperscripts \textquotedblleft $m$" and \textquotedblleft $s$"
respectively indicate maximality and symmetric nature. The fact that the
2-graded vector space $\mathfrak{so}_{9}\oplus \mathbf{16}$ can be endowed
with the structure of a (simple, exceptional) Lie algebra, and thus
satisfies the Jacobi identity (in particular, for three elements in $\mathbf{%
16}$), relies on a remarkable Fierz identity for $\mathfrak{so}_{9}$ gamma
matrices.

\item At the level of the \textit{reduced structure} Lie algebra $\mathfrak{%
str}_{0}$, one obtains the 3-graded structure of the real, minimally
non-compact form of $\mathfrak{e}_{6}$, namely :%
\begin{equation}
\mathfrak{str}_{0}\left( \mathbf{J}_{3}^{\mathbb{O}}\right) \supset ^{m,s}%
\mathfrak{str}_{0}\left( \mathbb{R}\oplus \mathbf{J}_{2}^{\mathbb{O}}\right)
\Leftrightarrow \left\{
\begin{array}{l}
\mathfrak{e}_{6(-26)}\supset ^{m,s}\mathfrak{so}_{9,1}\oplus \mathbb{R}; \\
~ \\
\begin{array}{c}
\mathfrak{e}_{6(-26)}=\mathbf{16}_{-1}^{\prime }\oplus \left( \mathfrak{so}%
_{9,1}\oplus \mathbb{R}\right) _{0}\oplus \mathbf{16}_{1}, \\
\text{or} \\
\mathfrak{e}_{6(-26)}=\mathbf{16}_{-1}\oplus \left( \mathfrak{so}%
_{9,1}\oplus \mathbb{R}\right) _{0}\oplus \mathbf{16}_{1}^{\prime },%
\end{array}%
\end{array}%
\right.  \label{e6}
\end{equation}%
where $\mathbf{16}$ and $\mathbf{16}^{\prime }$ are the Majorana-Weyl (MW)
spinors of $\mathbf{so}_{9,1}$, which constitute an example of \textit{%
Jordan pair} which is not a pair of Jordan algebras (see e.g. \cite{loos},
as well as \cite{pt1, pt2} for a recent treatment); also, the indeterminacy
denoted by \textquotedblleft or\textquotedblright\ depends on the \textit{%
spinor polarization} of the embedding \cite{Minchenko}. The fact that the
3-graded vector space(s) in the r.h.s. of (\ref{e6}) can be endowed with the
structure of a (simple, exceptional) Lie algebra, and thus satisfies the
Jacobi identity (in particular, for three elements in $\mathbf{16}\oplus
\mathbf{16}^{\prime }$), relies on a remarkable Fierz identity for $%
\mathfrak{so}_{9,1}$ gamma matrices. Note that $\mathfrak{str}\left( \mathbf{%
J}_{3}^{\mathbb{O}}\right) \simeq \mathfrak{str}_{0}\left( \mathbf{J}_{3}^{%
\mathbb{O}}\right) \oplus \mathbb{R}$ is isomorphic to the Lie algebra of
the automorphism group Aut$\left( \mathbf{J}_{3}^{\mathbb{O}},\mathbf{J}%
_{3}^{\mathbb{O}\prime }\right) $ of the \textit{Jordan pair} $\left(
\mathbf{J}_{3}^{\mathbb{O}},\mathbf{J}_{3}^{\mathbb{O}\prime }\right) $ :%
\begin{equation}
\mathfrak{str}\left( \mathbf{J}_{3}^{\mathbb{O}}\right) \simeq \text{Lie}%
\left( \text{Aut}\left( \left( \mathbf{J}_{3}^{\mathbb{O}},\mathbf{J}_{3}^{%
\mathbb{O}\prime }\right) \right) \right) \simeq \mathfrak{der}\left(
\mathbf{J}_{3}^{\mathbb{O}},\mathbf{J}_{3}^{\mathbb{O}\prime }\right) .
\end{equation}

\item At the level of the \textit{conformal} Lie algebra $\mathfrak{conf}$,
one obtains
\begin{equation}
\mathfrak{conf}\left( \mathbf{J}_{3}^{\mathbb{O}}\right) \supset ^{m,s}%
\mathfrak{conf}\left( \mathbb{R}\oplus \mathbf{J}_{2}^{\mathbb{O}}\right)
\Leftrightarrow \left\{
\begin{array}{l}
\mathfrak{e}_{7(-25)}\supset ^{m,s}\mathfrak{so}_{10,2}\oplus \mathfrak{sl}%
_{2,\mathbb{R}}; \\
~ \\
\mathfrak{e}_{7(-25)}=\mathfrak{so}_{10,2}\oplus \mathfrak{sl}_{2,\mathbb{R}%
}\oplus \left( \mathbf{32}^{(\prime )},\mathbf{2}\right) ,%
\end{array}%
\right.  \label{e7}
\end{equation}%
where $\mathbf{32}$ is the MW spinor of $\mathfrak{so}_{10,2}$, and the
possible priming (denoting spinor conjugation) depends on the choice of the
spinor polarization \cite{Minchenko}. By further branching the $\mathfrak{sl}%
_{2,\mathbb{R}}$, one obtain a 5-grading of contact type (recently
reconsidered within the classification worked out in \cite%
{Cantarini-Ricciardo-Santi}) of the real, minimally non-compact form of $%
\mathfrak{e}_{7}$, namely :%
\begin{equation}
\begin{array}{l}
\mathfrak{e}_{7(-25)}\supset \mathfrak{so}_{10,2}\oplus \mathbb{R}; \\
~ \\
\mathfrak{e}_{7(-25)}=\mathbf{1}_{-2}\oplus \mathbf{32}_{-1}^{(\prime
)}\oplus \left( \mathfrak{so}_{10,2}\oplus \mathbb{R}\right) _{0}\oplus
\mathbf{32}_{1}^{(\prime )}\oplus \mathbf{1}_{2}.%
\end{array}
\label{e7-2}
\end{equation}%
The fact that the 5-graded vector space(s) in the r.h.s. of (\ref{e7-2}) can
be endowed with the structure of a (simple, exceptional) Lie algebra, and
thus satisfies the Jacobi identity (in particular, for three elements in $%
\mathbf{32}^{(\prime )}\oplus \mathbf{32}^{(\prime )}$), relies on a
remarkable Fierz identity for $\mathfrak{so}_{10,2}$ gamma matrices. Note
that $\mathfrak{conf}\left( \mathbf{J}_{3}^{\mathbb{O}}\right) $ is
isomorphic to the Lie algebra of the automorphism group Aut$\left( \mathfrak{%
F}\left( \mathbf{J}_{3}^{\mathbb{O}}\right) \right) $ of the reduced
Freudenthal triple system constructed over $\mathbf{J}_{3}^{\mathbb{O}}$ :%
\begin{equation}
\mathfrak{conf}\left( \mathbf{J}_{3}^{\mathbb{O}}\right) \simeq \text{Lie}%
\left( \text{Aut}\left( \mathfrak{F}\left( \mathbf{J}_{3}^{\mathbb{O}%
}\right) \right) \right) \simeq \mathfrak{der}\left( \mathfrak{F}\left(
\mathbf{J}_{3}^{\mathbb{O}}\right) \right) .
\end{equation}

\item Finally, at the level of the \textit{quasi-conformal} Lie algebra%
\footnote{%
We recall that the quasi-conformal realization of $\mathfrak{e}_{8(-24)}$
concerns a non-linear action on an \textit{extended} derived Freudenthal
triple system $\mathfrak{EF}\left( \mathbf{J}_{3}^{\mathbb{O}}\right) \simeq
\mathbb{R}\oplus \mathfrak{F}\left( \mathbf{J}_{3}^{\mathbb{O}}\right) $
\cite{GKN}.} $\mathfrak{qconf}$ \cite{GKN, GP-04}, one obtains the 2-graded
structure of the real, minimally non-compact form of $\mathfrak{e}_{8}$,
namely :%
\begin{equation}
\mathfrak{qconf}\left( \mathbf{J}_{3}^{\mathbb{O}}\right) \supset ^{m,s}%
\mathfrak{qconf}\left( \mathbb{R}\oplus \mathbf{J}_{2}^{\mathbb{O}}\right)
\Leftrightarrow \left\{
\begin{array}{l}
\mathfrak{e}_{8(-24)}\supset ^{m,s}\mathfrak{so}_{12,4}; \\
~ \\
\mathfrak{e}_{8(-24)}=\mathfrak{so}_{12,4}\oplus \mathbf{128}^{(\prime )},%
\end{array}%
\right.  \label{e8}
\end{equation}%
where $\mathbf{128}$ is the MW spinor of $\mathfrak{so}_{12,4}$, and, again,
the possible priming (standing for spinorial conjugation) relates to the
choice of the spinor polarization \cite{Minchenko}. Further decomposition of
$\mathfrak{so}_{12,4}$ yields to a 5-grading of \textquotedblleft extended
Poincaré" type \cite{Cantarini-Ricciardo-Santi} :%
\begin{equation}
\begin{array}{l}
\mathfrak{e}_{8(-24)}\supset \mathfrak{so}_{11,3}\oplus \mathbb{R}; \\
~ \\
\mathfrak{e}_{8(-24)}=\left\{
\begin{array}{c}
\mathbf{14}_{-2}\oplus \mathbf{64}_{-1}^{\prime }\oplus \left( \mathfrak{so}%
_{11,3}\oplus \mathbb{R}\right) _{0}\oplus \mathbf{64}_{1}\oplus \mathbf{14}%
_{2}; \\
\text{or} \\
\mathbf{14}_{-2}\oplus \mathbf{64}_{-1}\oplus \left( \mathfrak{so}%
_{11,3}\oplus \mathbb{R}\right) _{0}\oplus \mathbf{64}_{1}^{\prime }\oplus
\mathbf{14}_{2},%
\end{array}%
\right.%
\end{array}
\label{e8-2}
\end{equation}%
where $\mathbf{64}$ is the MW spinor of $\mathfrak{so}_{11,3}$ and the
\textquotedblleft or\textquotedblright\ indeterminacy depends on the spinor
polarization \cite{Minchenko}. The fact that the 2-graded vector space $%
\mathfrak{so}_{12,4}\oplus \mathbf{128}^{(\prime )}$ can be endowed with the
structure of a (simple,exceptional) Lie algebra, and thus satisfies the
Jacobi identity (in particular, for three elements in $\mathbf{128}^{(\prime
)}$), relies on a remarkable Fierz identity for $\mathfrak{so}_{12,4}$ gamma
matrices. Equivalently, the fact that the 5-graded vector space(s) in the
r.h.s. of (\ref{e8-2}) can be endowed with the structure of a (simple,
exceptional) Lie algebra, and thus satisfies the Jacobi identity (in
particular, for three elements in $\mathbf{64}\oplus \mathbf{64}^{\prime }$%
), relies on a remarkable Fierz identity for $\mathfrak{so}_{11,3}$ gamma
matrices.
\end{enumerate}

\section{From Bott periodicity to Exceptional periodicity}

Thus, we have related the existence of (finite-dimensional, simple)
exceptional Lie algebras to some remarkable Fierz identities holding in $q+8$
dimensions (in particular, with signature $9+0$, $9+1$, $10+2$,and $12+4$,
for $q=1,2,4$ and $8$, respectively).

Now, by observing that the reality properties of spinors and the existence
and symmetry of invariant spinor bilinears are periodic \textit{mod }$8$
(Bott periodicity), one can define some algebras which (for the moment,
formally) generalize the spinor content of the real forms of exceptional Lie
algebras discussed above : these are the so-called \textit{\textquotedblleft
Exceptional Periodicity\textquotedblright } (EP)\textit{\ algebras }\cite%
{trm1,3bis}, and, as vector spaces, they are defined as follows ($n\in
\mathbb{N\cup }\left\{ 0\right\} $ throughout\footnote{%
Note that there has been a shift of unity with respect to the notation of
\cite{3bis} and \cite{trm1} : the index $n$ used here is actually $n-1$ of
such Refs..}) :

\begin{enumerate}
\item Level $\mathfrak{der}$ :
\begin{equation}
\mathfrak{f}_{4(-52)}^{(n)}:=\mathfrak{so}_{9+8n}\oplus \mathbf{\psi }_{%
\mathbf{so}_{9+8n}},  \label{EP-1}
\end{equation}%
where $\mathbf{\psi }_{\mathbf{so}_{9+8n}}\equiv \boldsymbol{2}^{4+4n}$~is
the Majorana~spinor~of~$\mathfrak{so}_{9+8n}.$

\item Level $\mathfrak{str}_{0}$ :%
\begin{equation}
\mathfrak{e}_{6(-26)}^{(n)}:=\mathbf{\psi }_{\mathbf{so}_{9+8n,1},-1}^{%
\prime }\oplus \left( \mathfrak{so}_{9+8n,1}\oplus \mathbb{R}\right)
_{0}\oplus \mathbf{\psi }_{\mathbf{so}_{9+8n,1},1},  \label{EP-2}
\end{equation}%
where $\mathbf{\psi }_{\mathbf{so}_{9+8n,1}}\equiv \boldsymbol{2}^{4+4n}$~is
the MW~spinor~of~$\mathfrak{so}_{9+8n,1}.$

\item Level $\mathfrak{conf}$ :%
\begin{eqnarray}
\mathfrak{e}_{7(-25)}^{(n)} &:&=\left( \mathfrak{so}_{10+8n,2}\oplus
\mathfrak{sl}_{2,\mathbb{R}}\right) \oplus \left( \mathbf{\psi }_{\mathbf{so}%
_{10+8n,2}},\mathbf{2}\right)  \label{EP-3} \\
&=&\mathbf{1}_{-2}\oplus \mathbf{\psi }_{\mathbf{so}_{10+8n,2},-1}\oplus
\left( \mathfrak{so}_{10+8n,2}\oplus \mathbb{R}\right) _{0}\oplus \mathbf{%
\psi }_{\mathbf{so}_{10+8n,2},1}\oplus \mathbf{1}_{2},  \notag
\end{eqnarray}%
where $\mathbf{\psi }_{\mathbf{so}_{10+8n,2}}\equiv \boldsymbol{2}^{5+4n}$%
~is the MW~spinor~of~$\mathfrak{so}_{10+8n,2}.$

\item Level $\mathfrak{qconf}$ :%
\begin{eqnarray}
\mathfrak{e}_{8(-24)}^{(n)} &:&=\mathfrak{so}_{12+8n,4}\oplus \mathbf{\psi }%
_{\mathbf{so}_{12+8n,4}}  \label{EP-4} \\
&=&\left( \mathbf{14+8n}\right) _{-2}\oplus \mathbf{\psi }_{\mathbf{so}%
_{11+8n,3},-1}^{\prime }\oplus \left( \mathfrak{so}_{11+8n,3}\oplus \mathbb{R%
}\right) _{0}\oplus \mathbf{\psi }_{\mathbf{so}_{11+8n,3},1}\oplus \left(
\mathbf{14+8n}\right) _{2},  \notag
\end{eqnarray}%
where $\mathbf{\psi }_{\mathbf{so}_{12+8n,4}}\equiv \boldsymbol{2}^{7+4n}$
and $\mathbf{\psi }_{\mathbf{so}_{11+8n,3}}\equiv \boldsymbol{2}^{6+4n}$
respectively denote the MW spinors of $\mathfrak{so}_{12+8n,4}$ and of $%
\mathfrak{so}_{11+8n,3}$.
\end{enumerate}

\begin{figure}[t]
\centering
\includegraphics[width=0.60\textwidth]{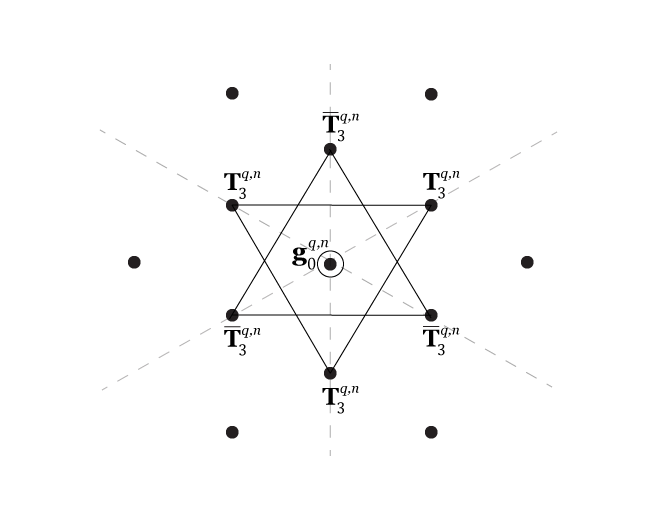}
\caption{The \textit{Magic Star} structure of the $\mathfrak{a}_{2}$%
-projection of the generalized root lattices of EP algebras.
finite-dimensional \protect\cite{trm1}. $\mathbf{T}_{3}^{q,n}$ stands for a
Vinberg T-algebra of rank-3 and of \textit{special type} \protect\cite%
{Vinberg}, parametrized by $q=1,2,4,8$ and $n\in N\cup \left\{ 0\right\} $,
corresponding to $\mathfrak{f}_{4}^{(n)}$, $\mathfrak{e}_{6}^{(n)}$, $%
\mathfrak{e}_{7}^{(n)}$, $\mathfrak{e}_{8}^{(n)}$, respectively.}
\label{fig:MagicStarT}
\end{figure}

A rigorous algebraic definition of the above EP algebras has been given in
\cite{trm1} (see also \cite{3bis}) by introducing the notion of \textit{%
generalized} roots, and by defining the structure constants in terms of (a
suitably generalized) Kac's \textit{asymmetry function} \cite{graaf,Kac}. In
this report, we confine ourselves to remark that EP algebras are \textit{not}
simply non-reductive nor semisimple, spinor-affine extensions of
(pseudo-)orthogonal Lie algebras, but their spinor generators are
non-translational (i.e., non-Abelian), as are the spinor generators of%
\footnote{%
The treatment on $\mathbb{R}$ given here is based on the EP generalization
of the various symmetry Lie algebras of the Albert algebra $\mathbf{J}_{3}^{%
\mathbb{O}}$, and it yielded to some specific real forms of $\mathfrak{f}%
_{4}^{(n)}$, $\mathfrak{e}_{6}^{(n)}$, $\mathfrak{e}_{7}^{(n)}$ and $%
\mathfrak{e}_{8}^{(n)}$. Starting from $\mathbb{C}$, a rigorous definition
of all real forms of EP algebras, by means of the introduction of suitable
involutive morphisms within the corresponding EP generalized root lattices
\cite{trm1}, will be the object of forthcoming works.} $\mathfrak{f}%
_{4(-52)}\equiv \mathfrak{f}_{4(-52)}^{(n=0)}$, $\mathfrak{e}_{6(-26)}\equiv
\mathfrak{e}_{6(-26)}^{(n=0)}$, $\mathfrak{e}_{7(-25)}\equiv \mathfrak{e}%
_{7(-25)}^{(n=0)}$, and $\mathfrak{e}_{8(-24)}\equiv \mathfrak{e}%
_{8(-24)}^{(n=0)}$. This yields to the \textit{violation} of the Jacobi
identity when considering three spinor generators as input in the Jacobiator
\cite{trm1}. As of today, a rigorous, axiomatic treatment of EP algebras is
missing : can EP algebras be defined in terms of some characterizing
identities, going beyond Jacobi? This remains an open problem.

The crucial result, which motivates and renders all the above construction
and the corresponding construction in the EP lattices non-trivial, is the
following \cite{trm1}: for $n>0$, all EP algebras admit a $\mathfrak{a}_{2}$
subalgebra, such that the projection of their generalized root lattices onto
the 2-dimensional plane defined by the Cartans of such $\mathfrak{a}_{2}$
has a\textit{\ Magic Star structure}, with those generalized roots
corresponding to the degeneracies on the tips of such EP-generalized Magic
Star which can be endowed with an algebraic structure, denoted by $\mathbf{T}%
_{3}^{q,n}$, generalizing the rank-3 simple Jordan algebras $\mathbf{J}%
_{3}^{q}\equiv \mathbf{J}_{3}^{\mathbb{A}}\equiv H_{3}\left( \mathbb{A}%
\right) $ mentioned above. The resulting, EP-generalized Magic Star is
depicted in Fig. \ref{fig:MagicStarT}. Remarkably, such a generalization is%
\footnote{%
Within a set of reasonable and intuitive assumptions \cite{Vinberg}.} the
\textit{unique} possible one, and it is provided by the Hermitian part of (a
class of) rank-3 T-algebras \textit{of special type}. Such algebras were
introduced some time ago by Vinberg \cite{Vinberg}, and they recently
appeared in \cite{AC, AMS1, AMS2}, in which they have been named \textit{%
Vinberg special T-algebras}.

\section{Vinberg special T-algebras and Bekenstein-Hawking entropy}

The real forms of EP algebras resulting from the treatment given above, i.e.
$\mathfrak{f}_{4(-52)}^{(n)}$, $\mathfrak{e}_{6(-26)}^{(n)}$, $\mathfrak{e}%
_{7(-25)}^{(n)}$, and $\mathfrak{e}_{8(-24)}^{(n)}$ (corresponding to $%
\mathfrak{der}$, $\mathfrak{str}_{0}$, $\mathfrak{conf}$ and $\mathfrak{qconf%
}$ levels, or, equivalently - by the symmetry of the Freudenthal-Tits Magic
Square \cite{tits1,freu} - to $q=1,2,4$ and $8$, respectively), the $3\times
3$ generalized matrix algebras $\mathbf{T}_{3}^{q,n}$ corresponding to the
set of generalized roots degenerating to a point on each of the tips of the
EP-generalized Magic Star (depicted in Fig. \ref{fig:MagicStarT}) can be
realized as follows :%
\begin{equation}
\mathbf{T}_{3}^{q,n}:=\left(
\begin{array}{ccc}
r_{1} & \mathbf{V}_{\mathfrak{so}_{q+8n}} & \mathbf{\psi }_{\mathfrak{so}%
_{q+8n}} \\
\overline{\mathbf{V}}_{\mathfrak{so}_{q+8n}} & r_{2} & \mathbf{\psi }_{%
\mathfrak{so}_{q+8n}}^{\prime } \\
\overline{\mathbf{\psi }}_{\mathfrak{so}_{q+8n}} & \overline{\mathbf{\psi }%
^{\prime }}_{\mathfrak{so}_{q+8n}} & r_{3}%
\end{array}%
\right) ,  \label{T}
\end{equation}%
where\footnote{$\left[ \cdot \right] $ denotes the integer part throughout.}%
\begin{eqnarray}
\mathbf{V}_{\mathfrak{so}_{q+8n}} &:&=\left( \boldsymbol{q+8n,1}\right) ; \\
\mathbf{\psi }_{\mathfrak{so}_{q+8n}} &:&=\left( \boldsymbol{2}^{\left[
\left( q+1\right) /2\right] +4n-1+\delta _{q,1}},\mathbf{Fund}\left(
\mathcal{S}_{q}\right) \right) ;
\end{eqnarray}
are irreducible representation spaces of the Lie algebra%
\begin{equation}
\mathfrak{so}_{q+8n}\oplus \mathcal{S}_{q},
\end{equation}%
with
\begin{equation}
\mathcal{S}_{q}:=\mathfrak{tri}_{\mathbb{A}}\ominus \mathfrak{so}_{\mathbb{A}%
}=0,\mathfrak{so}_{2},\mathfrak{su}_{2},0 ~\text{for~}%
q=1,2,4,8~\text{(i.e.,~for~}\mathbb{R}\text{,}\mathbb{C}\text{,}\mathbb{H}%
\text{,}\mathbb{O}\text{,~resp.)}
\end{equation}%
denoting the coset algebra of the \textit{triality} symmetry $\mathbf{tri}_{%
\mathbb{A}}$ of $\mathbb{A}$ \cite{Barton-Sudbery}:%
\begin{equation}
\begin{array}{lll}
\mathfrak{tri}_{\mathbb{A}}: & = & \left\{ \left( A,B,C\right)
|A(xy)=B(x)y+xC(y),~A,B,C\in \mathfrak{so}_{\mathbb{A}},~x,y\in \mathbb{A}%
\right\} \\
~ & ~ & ~ \\
~ & = & 0,\mathfrak{so}_{2}^{\oplus 2},\mathfrak{so}%
_{3}{}^{\oplus 3},\mathfrak{so}_{8}~\text{for~}\mathbb{A}=\mathbb{R},\mathbb{%
C},\mathbb{H},\mathbb{O}%
\end{array}%
\end{equation}%
modded by the \textit{norm-preserving} symmetry $\mathfrak{so}_{\mathbb{A}}$%
of $\mathbb{A}$ :%
\begin{equation}
\mathfrak{so}_{\mathbb{A}}:=\mathfrak{so}_{q}=0,\mathfrak{so}_{2},%
\mathfrak{so}_{4},\mathfrak{so}_{8}~\text{for~}\mathbb{A}=\mathbb{R},\mathbb{%
C},\mathbb{H},\mathbb{O}.
\end{equation}%
Actually, $\mathcal{S}_{q}$ is related to the reality properties of the
spinors of $\mathfrak{so}_{q+8n}$, and in Physics it is named $\mathcal{R}$%
-symmetry. Furthermore, $\mathbf{Fund}\left( \mathcal{S}_{q}\right) $
denotes the smallest non-trivial representation of the simple Lie algebra $%
\mathcal{S}_{q}$ (if any) :
\begin{equation}
\mathbf{Fund}\left( \mathcal{S}_{q}\right) =\mathbf{-},\mathbf{2},\mathbf{2}%
,-\mathbf{~}\text{for~}q=1,2,4,8,
\end{equation}%
with real dimension%
\begin{equation}
\text{fund}_{q}:=\dim _{\mathbb{R}}\mathbf{Fund}\left( \mathcal{S}%
_{q}\right) =1,2,2,1\mathbf{~}\text{for~}q=1,2,4,8.
\end{equation}%
Thus, the total real dimension of $\mathbf{T}_{3}^{q,n}$ is%
\begin{equation}
\dim _{\mathbb{R}}(\mathbf{T}_{3}^{q,n})=q+3+8n+\text{fund}_{q}\cdot 2^{%
\left[ (q+1)/2\right] +4n+\delta _{q,1}}.
\end{equation}

As mentioned above, $\mathbf{T}_{3}^{q,n}$ (\ref{T}) is the Hermitian part
of a certain class of generalized matrix algebras going under the name of
rank-3 T-algebras, introduced sometime ago by Vinberg as a unique,consistent
generalization of rank-3, simple Jordan algebras, within its theory of
homogeneous convex cones \cite{Vinberg} : more precisely, $\mathbf{T}%
_{3}^{q,n}$ has been dubbed \textit{exceptional }T-algebra in Sec. 4.3 of
\cite{3bis}. Upon a slight generalization (i.e., by including $P+\dot{P}$
copies of spinor irreprs., and correspondingly extending $\mathcal{S}_{q}$
to the \textquotedblleft full-fledged\textquotedblright\ $\mathcal{R}$%
-symmetry $\mathcal{S}_{q}\left( P,\dot{P}\right) $), $\mathbf{T}_{3}^{q,n}$
gets generalized to $\mathbf{T}_{3}^{q,n,P,\dot{P}}$ (with $P,\dot{P}\in
\mathbb{N}\cup \left\{ 0\right\} $), which occur in the study of so-called
homogeneous \textit{real special} manifolds\footnote{%
And, of course, in their images under R-map and c-map (cfr. e.g. \cite{dWVVP}%
, and Refs. therein).}. These are non-compact Riemannian spaces occurring as
(vector multiplets') scalar manifolds of $\mathcal{N}=2$-extended
Maxwell-Einstein supergravity theories in $D=4+1$ space-time dimensions,
firstly discussed to some extent by Cecotti \cite{Cecotti}. More recently, $%
\mathbf{T}_{3}^{q,n,P,\dot{P}}$ have appeared under the name of \textit{%
Vinberg special T- algebras} in works on Vinberg's theory of homogeneous
cones (and generalizations thereof) and on its relation to the entropy of
extremal black holes in $\mathcal{N}=2$-extended Maxwell-Einstein
supergravity theories in $D=3+1$ space-time dimensions \cite{AC,AMS1,AMS2}, .

The \textit{unique} invariant structure of the algebra\footnote{%
Correspondingly, $\mathcal{S}_{q}\equiv \left. \mathcal{S}_{q}(P,\dot{P}%
)\right\vert _{P=1,\dot{P}=0}$.} $\mathbf{T}_{3}^{q,n}\equiv \left. \mathbf{T%
}_{3}^{q,n,P,\dot{P}}\right\vert _{P=1,\dot{P}=0}$ given by (\ref{T}) is
provided by its (formal) \textquotedblleft determinant\textquotedblright .
In order to define it, let us introduce ($\mu =0,1,...,q+1+8n$)%
\begin{equation}
V^{\mu }:=\left( r_{1},r_{2},\mathbf{V}_{\mathbf{so}_{q+8n}}\right) ,
\label{V}
\end{equation}%
which, by recalling (\ref{lightcone}), is recognized to be a vector module
of $Spin\left( q+1+8n,1\right) $; we also denote the corresponding spinor of
$\mathfrak{so}_{q+1+8n,1}$ (which is chiral for $q=2,4,8$), of real
dimension fund$_{q}\cdot 2^{\left[ \left( q+1\right) /2\right] +4n+\delta
_{q,1}}$, by $\Psi ^{\alpha A}$ (where $\alpha =1,...,2^{\left[ \left(
q+1\right) /2\right] +4n+\delta _{q,1}}$ and $A=1,..,$fund$_{q}$). Then, the
\textquotedblleft determinant\textquotedblright\ of the generalized
Hermitian matrix algebra $\mathbf{T}_{3}^{q,n}$, which defines the cubic
norm $\mathbf{N}$ of $\mathbf{T}_{3}^{q,n}$ itself, is defined as
\begin{equation}
\mathbf{N}\left( \mathbf{T}_{3}^{q,n}\right) :=\frac{1}{2}\eta _{\mu \nu }%
\left[ r_{3}V^{\mu }V^{\nu }+\gamma _{\alpha \beta }^{\mu }\Psi ^{\alpha
A}\Psi _{A}^{\beta }V^{\nu }\right] ,
\end{equation}%
where $\eta _{\mu \nu }$ is the symmetric bilinear invariant of the vector
module $V$ (\ref{V}) of $Spin\left( q+1+8n,1\right) $, and $\gamma _{\alpha
\beta }^{\mu }$ are the gamma matrices of $\mathfrak{so}_{q+1+8n,1}$.

Remarkably, Ferrar's classification \cite{Ferrar} of elements of a rank-3
Jordan algebras in terms of \textit{invariant} \textit{rank}$=0,1,2,3$ can
be generalized to the classification of the elements of $\mathbf{T}%
_{3}^{q,n} $ depending on their invariant rank as well, defined as follows
\cite{trm1}:
\begin{equation}
\begin{array}{lll}
\text{rank-}3 & : & \mathbf{N}\neq 0; \\
\text{rank-}2 & : & \mathbf{N}=0; \\
\text{rank-}1 & : & \partial \mathbf{N}=0.%
\end{array}%
\end{equation}%
In those (ungauged) $\mathcal{N}=2$-extended Maxwell-Einstein supergravity
theories in $D=4+1$ space-time dimensions based on $\mathbf{T}_{3}^{q,n}$
\cite{Cecotti}, the magnetic charges of extremal black strings (with
near-horizon geometry $AdS_{3}\otimes S^{3}$) fit into $\mathbf{T}_{3}^{q,n}$
itself, and its Bekenstein-Hawking entropy $S_{BS}$ enjoys the interestingly
simple expression
\begin{equation}
S_{BS}=\pi \sqrt{\left\vert \mathbf{N}\right\vert }.
\end{equation}%
We conclude this report by pointing out that the entropy of the extremal
dyonic black holes in the corresponding (ungauged) $\left( 3+1\right) $%
-dimensional supergravity theory (obtained by compactifying the fourth
spacial dimension on $S^{1}$ and keeping the massless sector) has been
recently discussed in \cite{AMS1}. Analogue formulæ hold when considering
the most general case $\mathbf{T}_{3}^{q,n,P,\dot{P}}$ (with $P,\dot{P}\in
\mathbb{N}\cup \left\{ 0\right\} $).

\end{document}